\documentclass[slac_one]{revtex4}
\usepackage{graphicx}
\usepackage{fancyhdr}
\begin{document}

\title{{\small{2005 ALCPG \& ILC Workshops - Snowmass,
U.S.A.}}\\ 
\vspace{12pt}
Study of $V_LV_L\to t\bar{t}$ at the ILC Including 
${\cal O} (\alpha_s) $ QCD Corrections} 

%

\author{Stephen Godfrey}
\affiliation{Ottawa Carleton Institute for Physics, 
Department of Physics, Carleton University, Ottawa K1S 5B6 Canada}
\author{Shou-hua Zhu}
\affiliation{Institute of Theoretical Physics, School of Physics, 
Peking University, Beijing 100871, China}

\begin{abstract}
In the event that the Higgs mass is large or that the electroweak 
interactions are strongly interacting at high energy, top quark 
couplings to longitudinal components of the weak gauge bosons could 
offer important clues to the underlying dynamics.  
It has been suggested that 
precision measurements of $W_L W_L \to t\bar{t}$ and $Z_L Z_L \to t\bar{t}$ 
might provide hints of new physics.  In this paper we present results
for ${\cal O} (\alpha_s) $ QCD corrections to
$V_LV_L\to t\bar{t}$ scattering at the ILC.  We find that corrections to 
cross sections  can be as large as 30\% and must be 
accounted for in any precision measurement of $VV\to t\bar{t}$.

\end{abstract}

\maketitle

\thispagestyle{fancy}


\section{INTRODUCTION} 

Understanding the mechanism of electroweak symmetry breaking (EWSB) is a 
primary goal of the LHC and ILC \cite{Weiglein:2004hn}.  
While much effort has been devoted to the weakly interacting weak 
sector scenario the strongly interacting weak sector (SIWS) remains 
a possibility.  
Because the $t$-quark mass is the same order of magnitude as the scale of 
EWSB it has long been suspected that $t$-quark properties may provide 
hints about the nature of EWSB and the subprocess $V_LV_L\to t\bar{t}$ 
has been suggested as a probe.  
While $V_LV_L\to t\bar{t}$ can be studied at both hadron colliders and
$e^+e^-$ colliders the overwhelming QCD backgrounds will 
likely make it impossible to study 
the $V_LV_L\to t\bar{t}$ subprocess at the LHC \cite{Han:2003pu}.  

In contrast, the ILC offers a much cleaner environment.
The simplist 
approach is to study how the $t\bar{t}$ cross section varies with 
$M_H$ \cite{Gintner:1996cr}.
A more general
approach is to parametrize interactions in a nonlinearly realized 
electroweak chiral Lagrangian which is appropriate if the EWSB dynamics is 
strong with no Higgs bosons at low energies.  
A typical dimension five operator is 
$L^{eff}_1= (a_1 /\Lambda) \bar{t} t W_\mu^+ W^{-\mu}$,
where the coefficient $a_1$ is naively expected to be of order 1 when 
the cut-off of the theory is taken to be $\Lambda=4\pi v=3.1$~TeV. 
It is expected that $a_1$ can be measured to an 
accuracy of $\sim \pm 0.1$ at 95\% C.L.  \cite{Larios:1997ey}.
But to be able to attach meaning to precision 
measurements it is necessary to understand radiative corrections, both 
electroweak and QCD.  
In this contribution we present the results of a study of 
 ${\cal O} (\alpha_s) $ corrections to the tree level electroweak 
$V_L V_L \rightarrow t \bar t$ process in the SM at the ILC.
Due to space limitations we point the interested reader to 
Ref.~\cite{Godfrey:2004tj} for a more detailed account and a more 
complete set of references.

\section{CALCULATIONS AND RESULTS}

We are interested in the subprocesses $VV\to t\bar{t}$ which occur in 
the processes
$e^+e^- \to  \ell_1 \ell_2 +  V V \to \ell_1 \ell_2 + t\bar{t}$
where $\ell_1 \ell_2$ is $\nu\bar{\nu}$ for the $W^+W^-\to t\bar{t}$ 
subprocess and $e^+e^-$ for the $ZZ \to t\bar{t}$ subprocess. The 
vector bosons are treated as partons inside the $e^+$ and $e^-$ using 
the effective boson approximation \cite{eva,Dawson:1986tc}.  
The total cross section is then
obtained by integrating the $W$ (of $Z$) 
luminosities with the subprocess cross section \cite{Kauffman:1989aq}.

\begin{figure}[t]
\includegraphics[width=70mm]{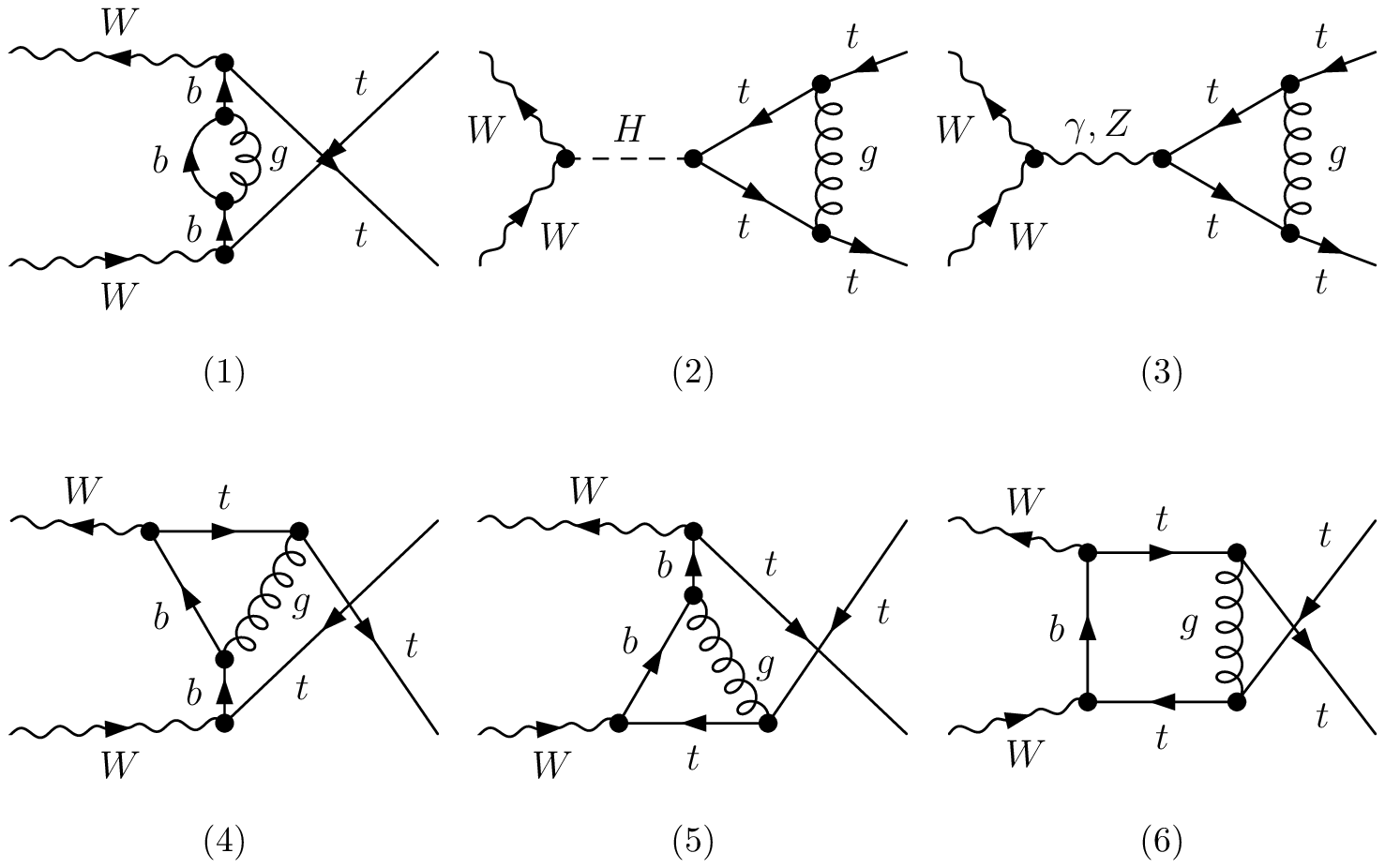}
\qquad\qquad
\includegraphics[width=70mm]{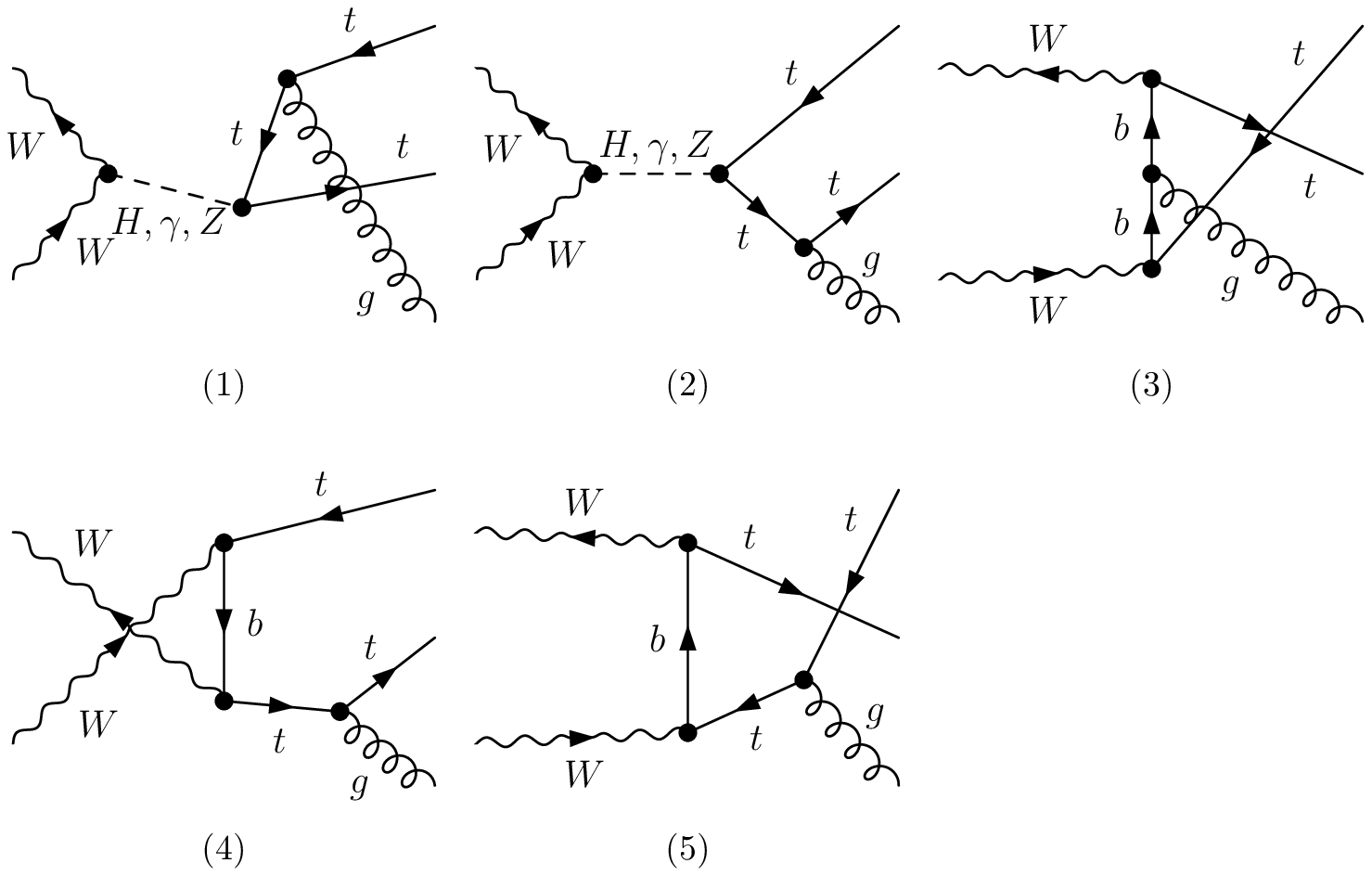}
\caption{${\cal O} (\alpha_s) $ QCD corrections to $W^+W^-\to t\bar{t}$.
(a) Virtual QCD contributions to $W^+W^-\to t\bar{t}$.
(b) Feynman diagrams for $W^+W^-\to t\bar{t}+g$. } \label{diagrams}
\end{figure}

The ${\cal O}(\alpha_s)$ corrections for 
the processes $W^+W^-\to t\bar{t}$ $ZZ\to t\bar{t}$ are calculated 
using the
FeynArts, FormCalc and LoopTools packages \cite{fclt}.  
The QCD corrections to $W^+W^- \to t\bar{t}$ are shown in 
Fig.~\ref{diagrams} (left side).
The infrared singularity in the vertex 
corrections are cancelled by the soft contributions from the process 
$W^+W^-\to t\bar{t}g$ which are shown in Fig.~\ref{diagrams} (right side). 
We regulate the IR-singularity by introducing a gluon mass which 
is equivalent to standard dimensional regularization
for processes with no triple gluon vertex present.
This approach has the additional benefit that
varying the value of the gluon mass acts as a check of 
the numerical cancellations between the different contributions.

The cross-sections are calculated by replacing the Born matrix 
element squared by 
\begin{equation}
|M_{Born}|^2 \to |M_{Born}|^2(1+\delta_{soft}) +2 {\rm Re}(M_{Born}^*\delta M)
\end{equation}
where $\delta M$ is the sum of the one-loop Feynman diagrams 
and the corresponding counter-term diagrams and 
$\delta_{soft}$ is the soft-gluon correction factor coming from the 
$2\to 3$ process.  

To deal with renormalization associated with the ultraviolet 
divergences we adopt the on-mass shell renormalization scheme and use
dimensional regularization.  
The software packages we used handle renormalization by
employing numerical 
factors.  Because our results must be independent of these factors 
we can vary them to check the consistency of our results. 
Likewise, we verified that the results are independent of the soft 
cutoff energy of the emitted gluon which divides the cross section 
into a piece with a soft gluon emitted and a piece with a hard gluon 
emitted.  We estimate uncertainties due to scale 
dependence results in at most a $\sim 4\%$ uncertainty in the K-factor.

\begin{figure}[b]
\includegraphics[width=70mm]{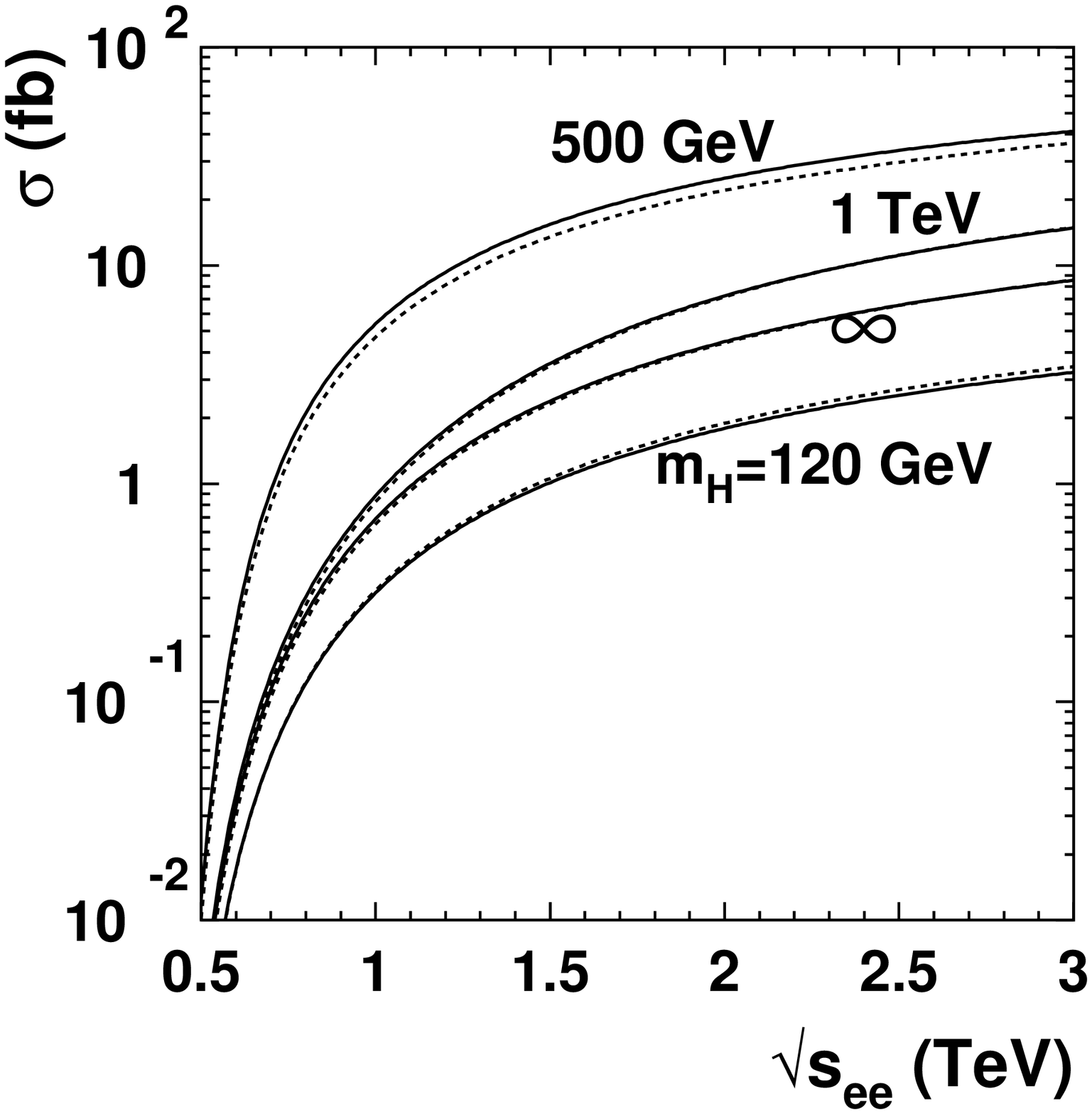}
\includegraphics[width=70mm]{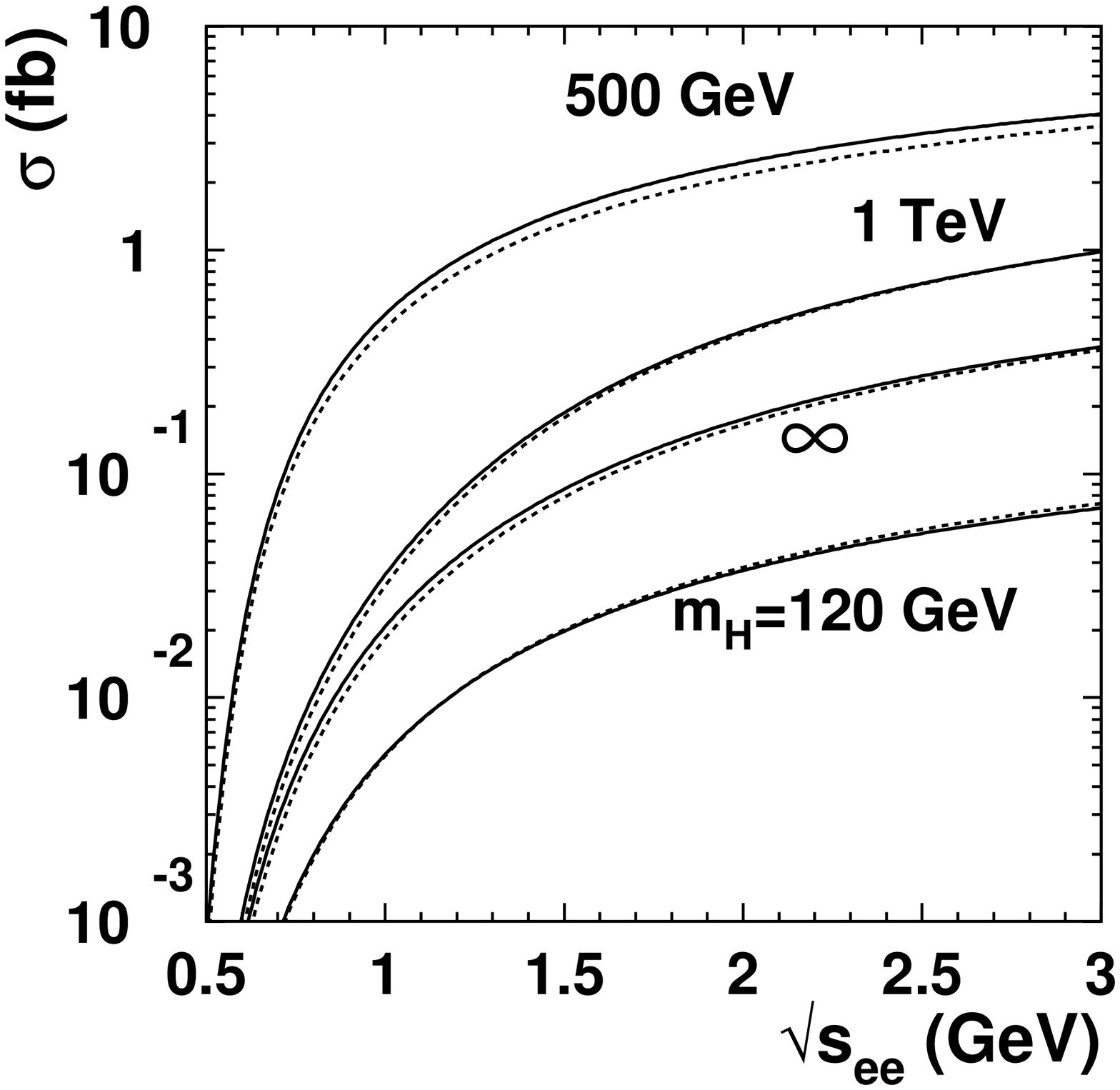}
\caption[]{
Cross sections as a function of $\sqrt{s}_{e^+e^-}$
for (a) $e^+e^- \to \nu\bar{\nu} t \bar t$ via $W^+_LW^-_L$ fusion 
and for (b) $e^+e^-\to e^+e^-  t \bar t$ via $Z_LZ_L$ fusion.
In both cases 
the solid line is the ${\cal O} (\alpha_s) $ 
QCD corrected cross section and the dashed line is the 
electroweak tree level
cross section. } \label{xsection}
\end{figure}

We include in our results the kinematic cuts
$m_{t\bar{t}}>400$~GeV and $p_T^{t,\bar{t}}>10$~GeV.  
Since the longitudinal scattering cross section is much larger 
than the $TT$ and $TL$ cases and it is the longitudinal gauge 
boson processes which corresponds to the Goldstone bosons of the 
theory we will henceforth only include results for $V_L V_L$ scattering. 
In Fig.~\ref{xsection} we show the 
cross section only including longitudinal $W$ and $Z$ 
scattering as a function 
of the $e^+e^-$ centre of mass energy for several representative Higgs 
masses including the $M_H\to \infty$ case (corresponding to the LET).

\begin{figure}[t]
\includegraphics[width=70mm]{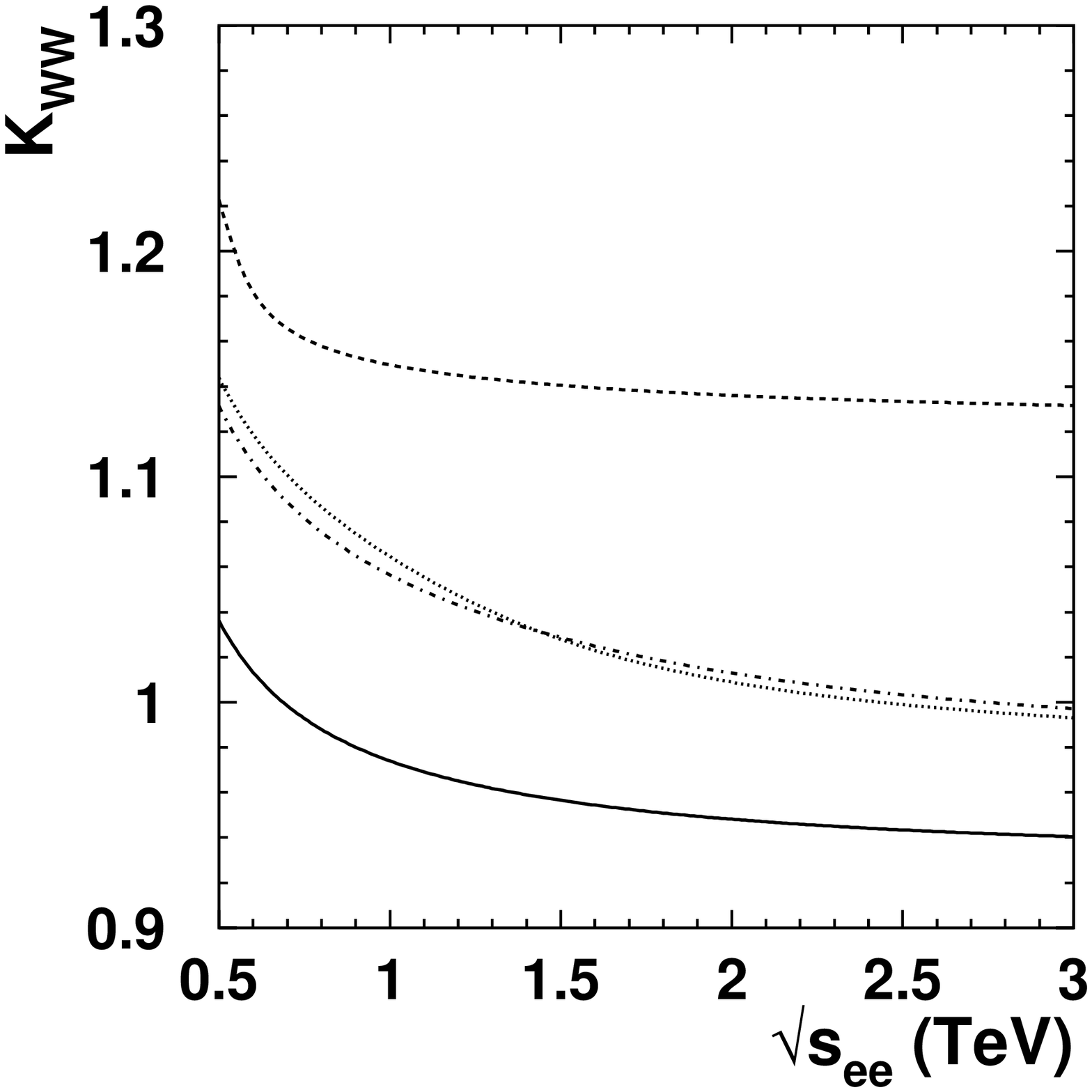}
\includegraphics[width=70mm]{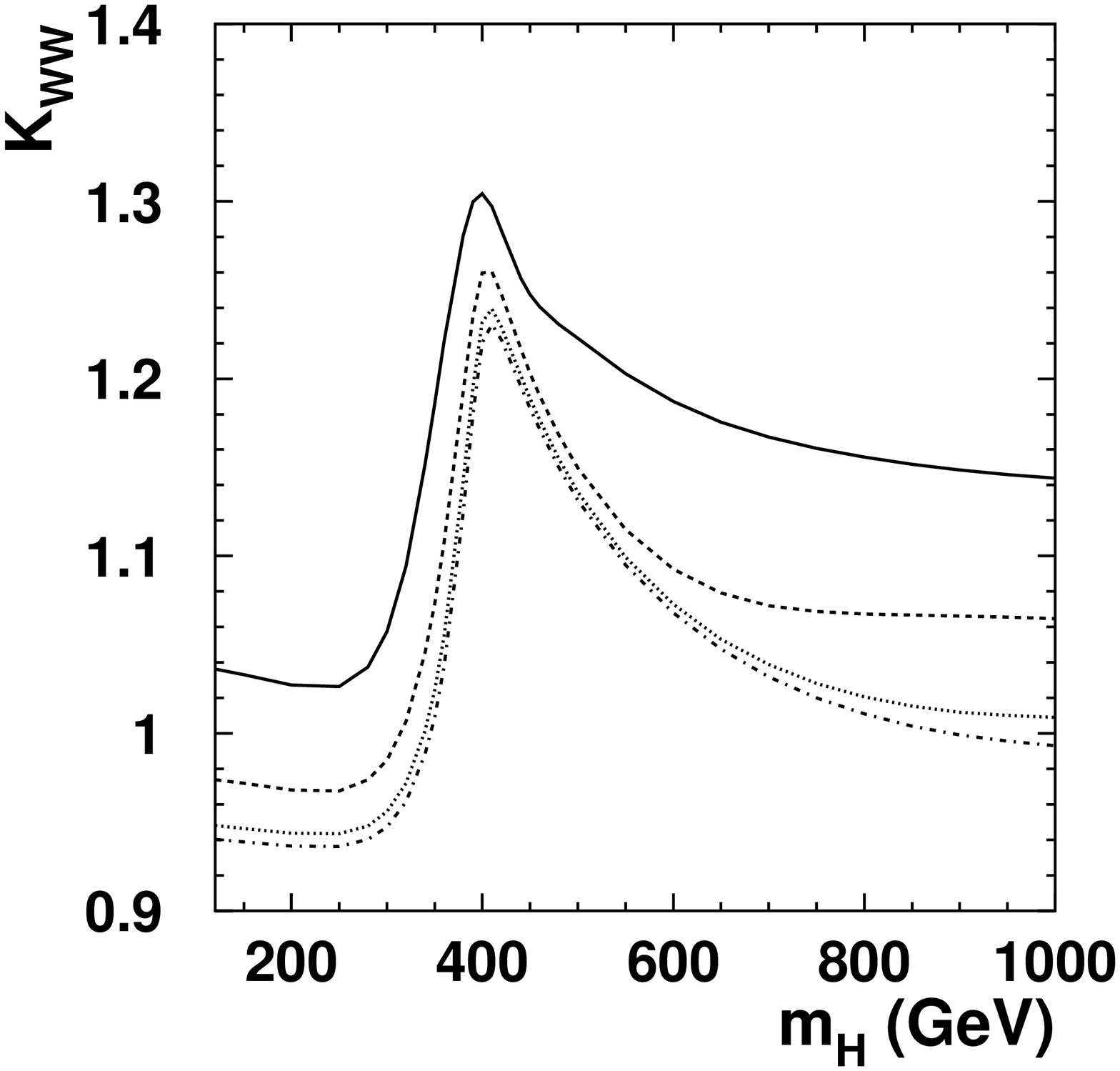}
\caption[]{(left side)
The K-factor as a function of  $\sqrt{s}_{e^+e^-}$
for $e^+e^- \to \nu\bar{\nu} t \bar t$ (via $W^+_LW^-_L$ fusion).
The solid line is for $M_H=120$~GeV, the dashed line for
$M_H=500$~GeV, the dotted line for $M_H=1$~TeV, and the dot-dashed 
line for $M_H=\infty$ (LET). 
(right side)
The K-factor as a function of  $M_H$
for $e^+e^- \to \nu\bar{\nu} t \bar t$ (via $W^+_LW^-_L$ fusion).
The solid line is for $\sqrt{s}_{e^+e^-}=500$~GeV,
the dashed line for $\sqrt{s}_{e^+e^-}=1$~TeV,
the dotted line for $\sqrt{s}_{e^+e^-}=2$~TeV, and
the dot-dashed line for $\sqrt{s}_{e^+e^-}=3$~TeV.
See text for an explanation of the K-factor.
} \label{kfactor}
\end{figure}

The QCD corrections to longitudinal scattering 
are often presented 
as a K-factor, normally defined as the ratio of the NLO to LO cross sections. 
Because the ${\cal O} (\alpha_s) $ 
QCD corrections we calculated are LO 
corrections to a tree level electroweak result we take the 
K-factor to be the ratio of the cross section with the 
${\cal O} (\alpha_s) $ QCD 
corrections and the tree level electroweak cross sections. 
The K-factors for 
$\sigma(e^+e^- \to \nu\bar{\nu} t \bar t)$ which goes via
$W^+_LW^-_L$ fusion and for 
$\sigma(e^+e^-\to e^+e^-  t \bar t)$ is
shown in Fig.~\ref{kfactor} (left side) 
as a function of $\sqrt{s}_{e^+e^-}$.
The ${\cal O} (\alpha_s) $ 
QCD corrections are largest for $M_H=500$~GeV with K-factors  
ranging from over 1.2 for $\sqrt{s}_{e^+e^-}=500$~GeV to 1.15 for 
$\sqrt{s}_{e^+e^-}=1$~TeV. The corrections decrease
as $\sqrt{s}_{e^+e^-}$ increases.
The variation of the K-factor with $M_H$ is shown in 
Fig.~\ref{kfactor} (right side).  
The fact that the K-factor is largest for 
$M_H=500$~GeV in Fig.~\ref{kfactor} (left side) 
and that it peaks at $M_H\simeq 400$~GeV in Fig.~\ref{kfactor} (right side)
is a threshold effect which
is an artifact of the kinematic cut we imposed on the $t\bar{t}$ 
invariant mass.  The important point 
is that the QCD corrections are not insignificant 
compared to the effects we might wish to study such as top Yukawa 
couplings or anomalous $VVt\bar{t}$ couplings.

\section{FINAL COMMENTS}

This paper concentrated on ${\cal O} (\alpha_s) $ QCD corrections.  
Other  equally 
important considerations and contributions that we have not discussed 
here are the issue of backgrounds, electroweak radiative 
corrections for top quark production, and the accuracy of the 
effective $W$ approximation.  These are discussed in 
Ref.~\cite{Godfrey:2004tj} along with relevant references.

In the event that the Higgs mass is heavy and the electroweak sector 
is strongly interacting it is quite possible that the underlying 
theory will manifest itself in the interactions between the top quark 
and longitudinal component of gauge bosons. In this paper we studied the 
${\cal O} (\alpha_s) $ QCD corrections to 
the tree level electroweak process $VV\to t\bar{t}$.  
We found that they can be quite substantial, 
the same size as the effects we wish to study, so that they 
need to be taken into account when studying $t\bar{t}$ 
production.

\begin{acknowledgments}
This research was supported in part by
the Natural Sciences and Engineering Research Council of Canada. 
SZ was also supported in part by the Natural Science Foundation of 
China under grant no. 90403004
\end{acknowledgments}

\end{document}